# Spatiotemporal symmetries and energy-momentum conservation in uniform spacetime metamaterials

Iñigo Liberal, Antonio Ganfornina-Andrades and J. Enrique Vázquez-Lozano

Department of Electrical, Electronic and Communications Engineering, Institute of Smart Cities (ISC), Public University of Navarre (UPNA), 31006 Pamplona, Spain. Inigo.liberal@unavarra.es

***Abstract –*** *Spacetime metamaterials (ST-MMs) are opening new regimes of light-matter interactions based on the breaking of temporal and spatial symmetries, as well as intriguing concepts associated with synthetic motion. In this work, we investigate the continuous spatiotemporal translation symmetry of ST-MMs with uniform modulation velocity. Using Noether's theorem, we demonstrate that such symmetry entails the conservation of the energy-momentum. We highlight how energy-momentum conservation imposes constraints on the range of allowed light-matter interactions within ST-MMs, as illustrated with examples of the collision of electromagnetic and modulation pulses. Furthermore, we discuss the similarities and differences between the conservation of energy-momentum and relativistic effects. We believe that our work provides a step forward in clarifying the fundamental theory underlying ST-MMs.*

1. Introduction

Recent experimental efforts on the ultra-fast optical modulation of the refractive index of materials[1–6], as well as the electronic control of metamaterial waveguides[7–9] and metasurfaces[10], are opening the pathway towards new regimes of light-matter interactions. In essence, the possibility of modulating the properties of optical systems at time-scales comparable to the frequency of operation have opened a shared multidisciplinary field that currently lies under the names of four-dimensional (4D) optics[11], time-varying media (TVM)[12] and/or spacetime metamaterials (ST-MMs)[13,14].

As it is usual in physics, much of the new phenomena can be traced back to the breaking of symmetries that are conventionally preserved. Noether's theorem[15,16] states that any continuous symmetry is associated with a conserved quantity and, therefore, breaking a symmetry removes the constraints imposed by the conservation of such quantity. For example, breaking the continuous temporal translation symmetry lifts the restrictions imposed by energy conservation, which enables novel forms of amplification and light emission. Examples include modified quantum vacuum amplification effects[17,18], thermal emission[19], frequency shifting with quantum noise management[20], as well as emission from localized quantum emitters[21] and free-electrons[22]. Similarly, breaking time-reversal symmetry lifts the restrictions imposed by reciprocity, which has enabled new technologies of magnet-less nonreciprocal devices[23].

Despite the interest of the broken symmetries, an equally relevant question is what symmetries remain intact, and what are the limitations imposed by them. For example, a purely time-modulated metamaterial preserves spatial translation symmetries. Therefore, the Minkowski[24,25] momentum of the electromagnetic field is conserved[26], imposing correlations between forward and backward waves. Similarly, a purely time-modulated metamaterial preserves rotational symmetries, but generally breaks duality and temporal symmetries, emphasizing the underlying differences between spin angular momentum, helicity and chirality[27].

In this work, we investigate the spatiotemporal symmetries of uniform space-time metamaterials (USTMs), i.e., metamaterials where the spatiotemporal modulation of the permittivity and permeability can be characterized by a uniform velocity $v$. As it will be shown, USTMs present a spatiotemporal symmetry with characteristic velocity $v$, which leads to the conservation of energy-momentum relations. This condition, while less restrictive than the independent conservation of energy and momentum, imposes limitations for the light-matter interactions that can take place within a USTM.

USTMs are a popular class of spacetime metamaterials, which have been proposed for new nonreciprocal amplifications schemes[28–30], generalized gratings and metasurfaces[31–34], quantum light emission analogue to Hawking radiation[35,36], Cerenkov-like radiation[37], frictional forces[38], or the observation of the Fresnel drag[39], just to name a few. Our work provides a fresh perspective into these many applications, and reveals a constraint hidden within them. Notice that, other generalizations of space-time metamaterials, e.g., recent proposals for accelerated-modulations[40], would not present the same constraints. Therefore, our work highlight the existence of reduced symmetries and conservation laws in ST-MMs, and the need to account for them in order to fully understand the opportunities of this exciting and growing field.

## 2. Spatiotemporal translation symmetry

As schematically depicted in Fig. 1, we consider a USTM with a spatiotemporal modulation travelling along the $Z$-axis with uniform velocity $v$, characterized by permittivity $\varepsilon(z - vt)$ permeability $\mu(z - vt)$. We work in the Coulomb gauge[16], so that the electric $\boldsymbol{E}(\mathbf{r}, t) = -\partial_t \boldsymbol{A}(\mathbf{r}, t)$ and magnetic flux $\boldsymbol{B}(\mathbf{r}, t) = \nabla \times \boldsymbol{A}(\mathbf{r}, t)$ fields are determined by the vector potential $\boldsymbol{A}(\mathbf{r}, t)$. Since we are interested on the interaction with modes propagating along the direction of propagation of the modulation, without any loss of generality we consider a reduced one dimensional problem $\boldsymbol{A}(\mathbf{r}, t) = \hat{\mathbf{x}} A_x(z, t)$.

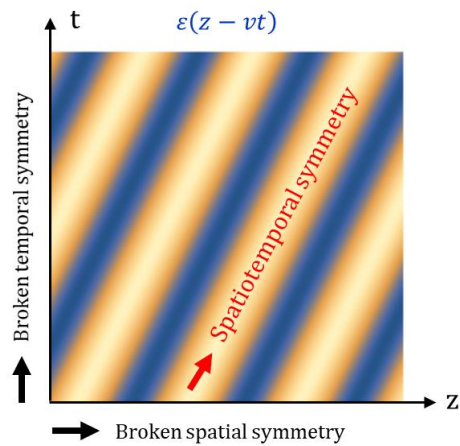

**Fig. 1. Spatiotemporal symmetries of uniform spacetime metamaterials (USTM).** Schematic depiction of a USTM where the permittivity $\varepsilon(z - vt)$ and/or the permeability $\mu(z - vt)$ presents a spatiotemporal modulation with characteristic velocity $v$. Such system has broken symmetries along the spatial and temporal axes, but still presents a continuous translation symmetry along the $z/t = v$ direction.

In this manner, Maxwell equations reduce to the following wave equation for the vector potential:

$$\partial_z \left\{ \frac{1}{\mu(z,t)} \partial_z A_x(z,t) \right\} = \partial_t \{\varepsilon(z,t) \partial_t A_x(z,t)\} \tag{1}$$

Then, we define a spatiotemporal continuous transformation consisting of a differential variation of the fields including temporal and spatial displacements weighted by a characteristic velocity $v$:

$$dA_x(z,t) = dt\, (\partial_t + v\partial_z)\, A_x(z,t) \tag{2}$$

In passing, we note that $(\partial_t + v\partial_z) A_x(z,t) = -E_x(z,t) + vB_y(z,t)$, a quantity whose continuity across a sharp spatiotemporal interfaces for moving modulations[13,14,41] and moving matter[42] is used as a boundary condition. Here, we remark that such quantity is not generally preserved for all times and points in space, but it represents a spatiotemporal displacement of the canonical variable $A_x(z,t)$, which serves for the investigation of the symmetries and globally conserved quantities of the system.

A continuous transformation corresponds with a symmetry when measurable outcomes are not changed by the transformation. As schematically depicted in Fig. 2, it implies that the dynamical variable $A_x(z,t)$ and its transformed counterpart $A'_x(z,t) = A_x(z,t) + dA_x(z,t)$ will experience exactly the same trajectories. Therefore, we first check if such continuous transformation is a symmetry of Maxwell equations, i.e., if $A'_x(z,t)$ also obeys the wave equation given by Eq. (1).

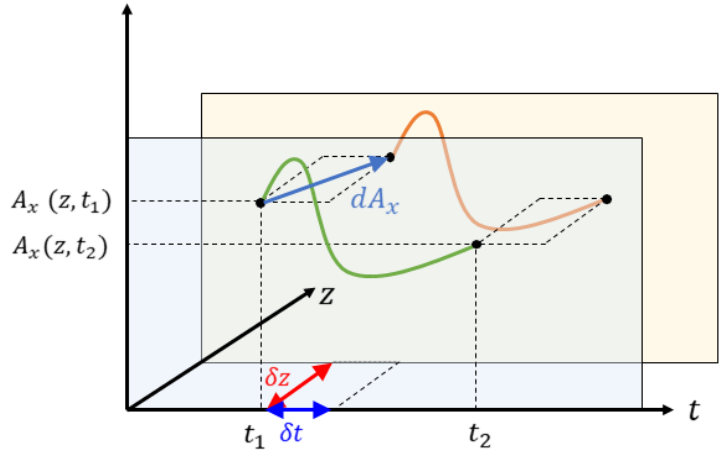

**Fig. 2. Trajectories of the vector potential under spatiotemporal translations.** Schematic depiction of how dynamical variable $A_x(z,t)$ exhibits the same trajectory under a continuous spatiotemporal translation, identifying it as a symmetry of the system.

As detailed in the Supplementary Material Section 1, by introducing $A'_x(z,t) = A_x(z,t) + dA_x(z,t)$ into the l.h.s. of (1) and operating, the following result can be derived

$$\partial_z \left\{ \frac{1}{\mu(z,t)} \partial_z A'_x(z,t) \right\} = \partial_t \{\varepsilon(z,t) \partial_t A'_x(z,t)\}$$

$$+ dt\{(\partial_t + v\partial_z)\{\mu(z,t)\varepsilon(z,t)\}\partial^2_t A_x(z,t)$$

$$+ (\partial_t + v\partial_z)\{\mu(z,t)\partial_t \varepsilon(z,t)\}\partial_t A_x(z,t)$$

$$+ (\partial_t + v\partial_z)\left\{ \frac{\mu(z,t)}{\partial_z \mu(z,t)} \right\} \partial_z A_x(z,t) \right\}$$

(3)

The existence of terms in the second, third and fourth rows of Eq. (3) reveal that, in general, the spatiotemporal continuous transformation (2) is not a symmetry of Maxwell equations for arbitrary spacetime modulations. However, it can be readily check that all these terms vanish provided that the permittivity and permeability modulations obey the following relations

$$(\partial_t + v\partial_z)\, \varepsilon(z,t) = 0$$

$$(\partial_t + v\partial_z)\, \mu(z,t) = 0$$

(4)

In other words, the continuous transformation (2) is a symmetry for a USTM with $\varepsilon(z,t) = \varepsilon(z - vt)$ and/or $\mu(z,t) = \mu(z - vt)$. Here, it is important to remark that a homogeneous permittivity $\varepsilon(z,t) = \varepsilon_{cst}$ and/or permeability $\mu(z,t) = \mu_{cst}$ is also a uniform spatiotemporal modulation for all velocities. In other words, homogeneous media is invariant under any spatiotemporal translations, while USTM is only invariant for a very specific spatiotemporal translation. Therefore, the conditions given by Eq. (4) are satisfied in four different scenarios (i) $\varepsilon(z - vt)$ and $\mu(z - vt)$, (ii) $\varepsilon(z - vt)$ and $\mu_{cst}$, (iii) $\varepsilon_{cst}$ and $\mu(z - vt)$, and (iv) $\varepsilon_{cst}$ and $\mu_{cst}$.

### 3. Noether's theorem and energy-momentum conservation

Next, we derive the conserved quantity associated with the spatiotemporal translation symmetry, finding that it is a linear superposition of the energy and the Minkowski momentum of the electromagnetic field. To this end, we start with a Lagrangian description of the electromagnetic field[16] $L(t) = \int dz\, \mathcal{L}(z,t)$, with Lagrangian density $\mathcal{L}(z,t) = \frac{\varepsilon(z,t)}{2}\left(\partial_t A_x(z,t)\right)^2 - \frac{1}{2\mu(z,t)}\left(\partial_z A_x(z,t)\right)^2$, where it can be checked that Euler-Lagrange's equation: $\frac{\partial \mathcal{L}}{\partial A_x} = \partial_t \frac{\partial \mathcal{L}}{\partial(\partial_t A_x)} + \partial_z \frac{\partial \mathcal{L}}{\partial(\partial_z A_x)}$ results in the wave equation (1).

Noether's theorem, as previously applied to time-varying media[26,27], states that for any symmetry of the system, $dA_x$, the following quantity must be a conserved quantity

$$\psi(dA_x) = \int dz\, \frac{\partial \mathcal{L}}{\partial(\partial_t A_x)} dA_x - L dt$$

(5)

Accordingly, for the spatiotemporal translation symmetry given by Eq. (2), the conserved quantity is given by

$$\psi(dA_x) = H(t) - vP(t)$$

(6)

where $H(t)$ is the electromagnetic energy

$$H(t) = \int dz \; \frac{\varepsilon(z,t)}{2} \left(\partial_t A_x(z,t)\right)^2 + \frac{1}{2\mu(z,t)} \left(\partial_z A_x(z,t)\right)^2 \tag{7}$$

and $P(t) = \int dz \; \hat{\mathbf{z}} \cdot (\mathbf{D} \times \mathbf{B})$ is the projection of the Minkowski[24,25] momentum of the electromagnetic field along $\hat{\mathbf{z}}$:

$$P(t) = -\int dz \; \varepsilon(z,t) \partial_t A_x(z,t) \partial_z A_x(z,t) \tag{8}$$

Therefore, the conserved quantity associated with a spatiotemporal translation is a linear superposition of energy and momentum. This point highlights that spatiotemporal metamaterials can have reduced symmetries, leading to less strict conservation laws. That is to say, energy and momentum are not required to be conserved individually, but there is a precise linear combination of them that does need to be conserved.

The coefficient of such linear superposition is given by the velocity of the modulation $v$. Consequently, the character of the energy-momentum quantity critically depends on $v$. In the $v \to 0$ limit, the spatiotemporal modulation converges to a purely spatial modulation, e.g., $(z - vt) \to \varepsilon(z)$, and the energy-momentum conservation converges to the conservation of energy $H(t) - vP(t) \to H(t)$. Inversely, in the $v \to \infty$ limit the energy-momentum approaches the Minkowski momentum, as it would be the case in a purely temporal modulation. Finally, energy and momentum are individually conserved in a homogeneous medium with neither spatial nor temporal modulations, as the $H(t) - vP(t)$ must be conserved for all $v$.

### 4. Continuity equations

A different perspective on the energy-momentum conservation can be found by deriving relevant continuity equations. To this end, we define energy $H(t) = \int dz \; h(z,t)$ and momentum $P(t) = \int dz \; p(z,t)$ densities. Then, by taking their time derivatives and rearranging the terms (see Supplementary Material Section 2), we find the following continuity equations:

$$\frac{dh(z,t)}{dt} + \frac{dF_h(z,t)}{dz} = J_h(z,t) \tag{9}$$

and

$$\frac{dp(z,t)}{dt} + \frac{dF_p(z,t)}{dz} = J_p(z,t) \tag{10}$$

where the energy flux is given by the Poynting vector: $F_h(z,t) = p(z,t)c^2(z,t)$, while the momentum flux is given by the energy density $F_p(z,t) = h(z,t)$. In addition, the source/sink term on the r.h.s. of Eq. (9) is given by

$$J_h(z,t) = \frac{1}{2}\left[-\partial_t \varepsilon(z,t)\left(\partial_t A_x(z,t)\right)^2 + \partial_t \mu^{-1}(z,t)\left(\partial_z A_x(z,t)\right)^2\right] \tag{11}$$

where one intuitively confirms that the source/sinks of the energy arise from the temporal derivatives of the material parameters: $\partial_t \varepsilon(z,t)$ and $\partial_t \mu^{-1}(z,t)$. On the other hand, the source/sink term on the r.h.s. of Eq. (10) is given by

$$J_p(z,t) = \frac{1}{2}\left[\partial_z \varepsilon(z,t)\big(\partial_t A_x(z,t)\big)^2 - \partial_z \mu^{-1}(z,t)\big(\partial_z A_x(z,t)\big)^2\right] \quad (12)$$

In this case, one intuitively confirms that the source/sinks of the momentum arise from the spatial derivatives of the material parameters: $\partial_z \varepsilon(z,t)$ and $\partial_z \mu^{-1}(z,t)$.

Therefore, the source/sink terms for the energy momentum density all comprise terms of the form $\partial_t \varepsilon(z,t) + v\partial_z \varepsilon(z,t)$ and $\partial_t \mu^{-1}(z,t) + v\partial_z \mu^{-1}(z,t)$ which cancel out for uniformly moving modulations $\varepsilon(z - vt)$ and/or $\mu(z - vt)$. From this perspective, the conservation of the energy-momentum stems from the fact that the source/sinks of energy and momentum locally cancel each other for uniformly moving modulations.

### 5. Differences with the four-momentum

Spatiotemporal metamaterials are often discussed as analogues of relativity effects and/or synthetic motion[14]. From this perspective, the structure of the conserved energy-momentum $H(t) - vP(t)$ presents some similarities with the four-momentum of a particle $\widetilde{P} = \left(\frac{E}{c}, \widetilde{P}_x, \widetilde{P}_y, \widetilde{P}_z\right)$ with conserved length $\widetilde{P} \cdot \widetilde{P} = -\frac{E^2}{c^2} + \widetilde{P}_x^2 + \widetilde{P}_y^2 + \widetilde{P}_z^2$, or the stress tensor of the electromagnetic field. The similarity arises from the mixing of energy and momentum through a preferred velocity, as well as the minus signed involved in the conservation. At the same time, it should be very clear that both conserved quantities emerge from different considerations and are fundamentally different. Examples of the differences between both quantities include: (i) The conserved value is a linear combination of energy and momentum, instead of a quadratic one. (ii) The scaling relationship between energy and momentum is given by the modulation velocity, instead of the velocity of light. (iii) For each modulation velocity, there is a different value of the conserved quantity $H - vP$, while $\widetilde{P} \cdot \widetilde{P}$ has the same value for all inertial frames, irrespectively of their associated velocity.

### 6. Consequences of the energy-momentum conservation

The conservation of the energy-momentum imposes constraints on the range of light-matter interactions that can take place within the framework of USTMs. Specifically, it imposes that the changes of energy and momentum must be intertwined. To illustrate this point, we note that $H(t_2) - vP(t_2) = H(t_1) - vP(t_1)$ can be rewritten as $\Delta H / \Delta P = v$, showing the explicit relation between energy and momentum variations mediated by the velocity of the modulation. A crucial factor is also the alignment of the direction of the modulation and the initial momentum of the electromagnetic field, i.e., if the electromagnetic fields and the modulation are copropagating or counterpropagating.

For example, let us consider the case in which the initial momentum and the velocity of the modulation are positive: $P(t_1) > 0$ and $v > 0$. In this case, "amplifying" in the sense of $\Delta H > 0$ implies that the momemtum must be increased by $\Delta H/v$. Furthermore, the faster the modulation the smaller the increase of momentum per unit of energy. In the $v \to \infty$ limit the momentum is conserved even if the energy changes. Conversely, in the $v \to 0$ limit energy is conserved, so that it does not change even if the momentum changes. However, "reflecting" in the sense of reverting the direction of the momentum of the field, i.e., $P(t_2) < 0$, or, equivalently, $\Delta P < -P(t_1)$, implies that the energy must be reduced by $\Delta H < -vP(t_1)$. Thus, it can only be achieved if there is enough initial energy $H(0) > vP(0)$.

On the other hand, in the counter-propagating case the initial momentum is positive $P(t_1) > 0$ and the velocity of the modulation is negative $v < 0$, then $\Delta H > 0$ implies that the momemtum must be decrased by $\Delta H/v$, with similar asymptotic limits. Furthermore, a decrease of the momentum is accompanied by amplification.

7. **Example of scattering interactions**

In this section, we utilize the theory above to analyze scattering interactions between localized electromagnetic pulses and travelling modulations. We use this scenario as a case study to illustrate different regimes for the exchange of energy and momentum, as well as the overall energy-momentum conservation. Specifically, we use a "Sech" electromagnetic pulse[43] defined via its initial conditions at $t = 0$, before any interaction takes place: $D_x(z, t = 0) = \text{Sech}(z/\Delta z)\sin(2\pi z/Z_p)$. We also use a travelling step modulation, i.e., $\varepsilon(z - vt) = \varepsilon_b f(z - vt)$ and $\mu(z - vt) = \mu_b f(z - vt)$ with $f(Z) = f_1$ for $Z \leq Z_1$ and $Z \geq Z_2$, and $f(Z) = f_2$ for $Z_1 < Z < Z_2$. This impedance-matched modulation results in a local modulation of the speed of light $c(z - vt) = c_b/f(z - vt)$ and $c_b = 1/\sqrt{\mu_b \varepsilon_b}$. It simplifies the analysis while keeping relevant features of the energy-momentum conservation, and it is a common modulation within the context of USTMs[29,35,39,44]. For such impedance-matched modulation, the partial differential equations that describe wave propagation can be solved via the method of characteristics[45]. In particular, the displacement field can be written as follows

$$D_x(z,t) = \frac{c(Z_0) - v}{c(z - vt) - v} D_x(Z_0, t) \tag{12}$$

where $Z_0$ is the crossing the with the $T = 0$ axis of the characteristic trayectory defined by $\frac{dT}{dZ} = \frac{1}{c(Z) - v}$, in the transformed coordinate system $Z = z - vt$ and $T = t$. All mathematical derivations involved are reported in Supplementary Material Section 3.

Our choice of examples target four different regimes: (i) Superluminal copropagation ($v = 2c_b$), (ii) subluminal copropagation ($v = c_b/2$), (iii) subluminal counterpropagation ($v = -c_b/2$), and (iv) superluminal counterpropagation in ($v = -2c_b$). The results are gathered together in Figs. 3 and 4. In addition, videos of the time evolution of the interactions for each example are included in Supplementary Material. The following set of parameters: $f_1 = 1$, $f_2 = 1.5$, $Z_2 - Z_1 = 10Z_p$, $\Delta z = 1.25Z_p$ ($Z_p = 1$) were used for all calculations.

First, in the case of <u>superluminal copropagation ($v = 2c_b$)</u>, the modulation pulse moves faster than the electromagnetic pulse, and the interaction takes place as the modulation passes through the electromagnetic pulse, which results in its compression (see Fig. 3a). In this case, both the energy and momentum of the electromagnetic pulse are increased, while the energy-momentum is confirmed to be conserved (see Fig. 4a). Moreover, the changes in energy are more prominent than those of the momentum, as expected from the superluminal regime. As $v$ gets close the luminal regime, the overlapping time between the electromagnetic and modulation pulses increases, and the increase of energy and momentum are more pronounced (see Supplementary Figure S1).

A qualitatively different scenario is observed in the case of subluminal copropagation ($v = c_b/2$). For this configuration, the modulation is slower than the electromagnetic pulse, and the interaction takes place as the electromagnetic pulse passes through the modulation (see Fig. 3b). Another difference is that the mode becomes expanded instead of being compressed. Moreover, the interaction with the modulation results in a decrease of both energy and momentum, while the energy-momentum is conserved (see Fig. 4b). Our results also confirm that the changes in the momentum become dominant as the modulation velocity is decreased. In fact, in the exact zero velocity case ($v = 0$), the modulation becomes purely spatial and energy is conserved while the momentum is not (see Supplementary Figure S2).

Yet a different regime appears in the case of subluminal counter-propagation ($v = -c_b/2$). Within this regime, the electromagnetic pulse and the modulation collision against each other (see Fig. 3c), resulting in an increase of the energy but a reduction of the momentum (see Fig. 4c). A similar behavior is observed for the case of superluminal counterpropagation ($v = 2c_b$), where the main difference is that the interaction takes place in a much shorter time scale (see Fig. 3d), and it becomes more energy-like, as predicted by the conservation of the energy-momentum at high modulation speeds (see Fig. 4d). Therefore, the transition from subluminal to superluminal does not present a qualitative change in the interaction, by contrast with the copropagating case. From a mathematical standpoint, this difference can be justified by the fact that $c(z - vt) - v$ in Eq. (12) does not change sign in the counter-propagating case. In the deeply superluminal regime, the interaction results in a short impulse of energy, while the momentum remains constant, in agreement with the convergence with pure time-varying media (see Supplementary Figure S3). To finalize, we remark that all the features observed in this section are consistent with the constraints discussed in the previous sections.

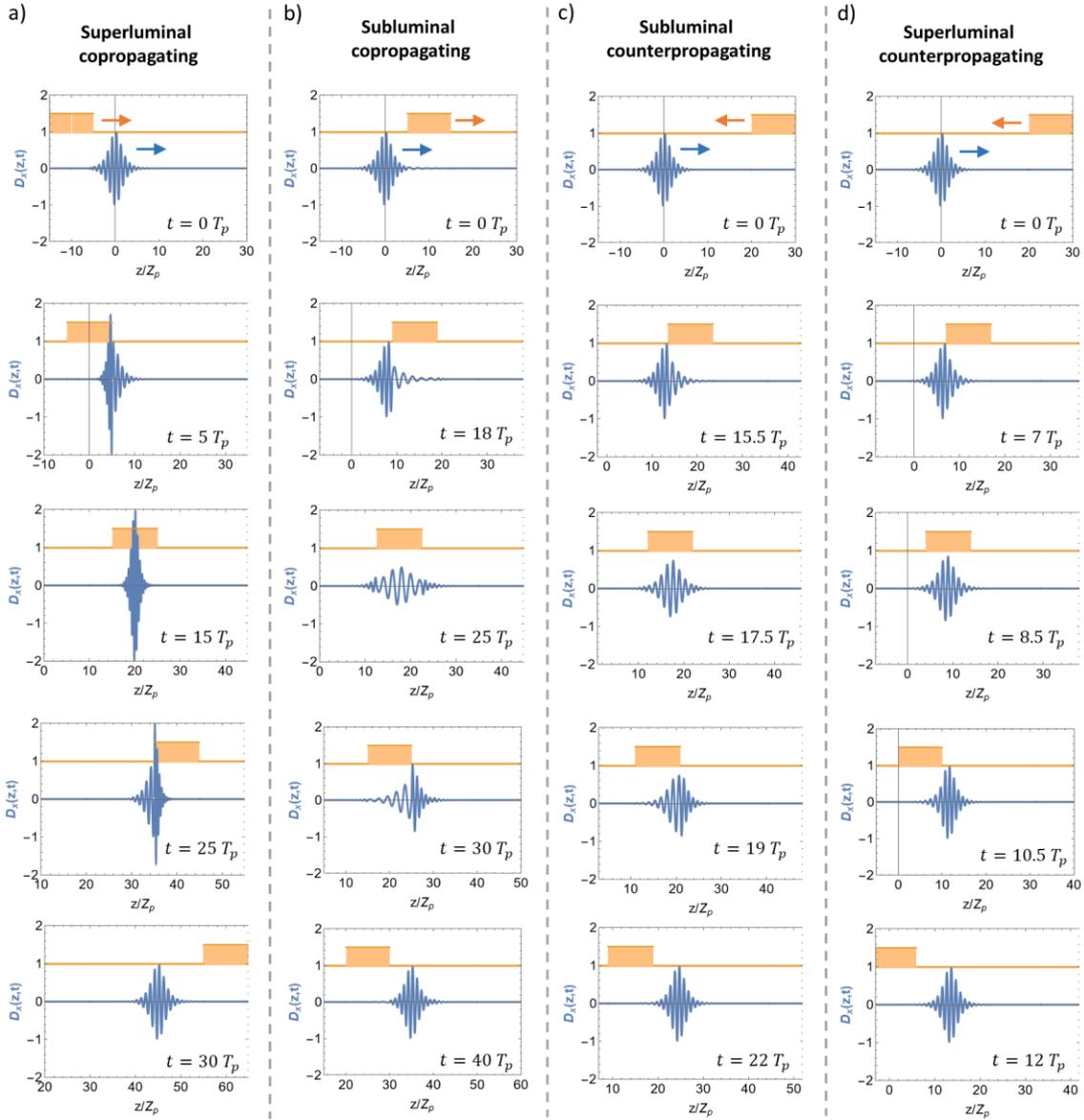

**Fig. 3. Snapshots of scattering interactions in different regimes.** Displacement field $D_x(z,t)$ (blue) and phase velocity $c(z-vt)$ (orange). (a) <u>Superluminal copropagating</u>: modulation velocity $v = 2c_b$, initial modulation position $Z_1/Z_p = -15$ and $Z_2/Z_p = -5$, snapshot times: $t/T_p = 0, 5, 15, 25, 30$. (b) <u>Subluminal copropagating</u>: modulation velocity $v = c_b/2$, initial modulation position $Z_1/Z_p = 5$ and $Z_2/Z_p = 15$, snapshot times: $t/T_p = 0, 18, 25, 30, 40$. (c) <u>Subluminal counterpropagating</u>: modulation velocity $v = -c_b/2$, initial modulation position $Z_1/Z_p = 20$ and $Z_2/Z_p = 30$, snapshot times: $t/T_p = 0, 15.5, 17.5, 19, 22$. (d) <u>Superluminal counterpropagating</u>: modulation velocity $v = -2c_b$, initial modulation position $Z_1/Z_p = 20$ and $Z_2/Z_p = 30$, snapshot times: $t/T_p = 0, 7, 8.5, 10.5, 12$. The following set of parameters: $f_1 = 1$, $f_2 = 1.5$, $Z_2 - Z_1 = 10Z_p$, $\Delta z = 1.25Z_p$, $T_p = Z_p/c_b$ were used in all four regimes. Videos associated with these snapshots are included as supplementary material.

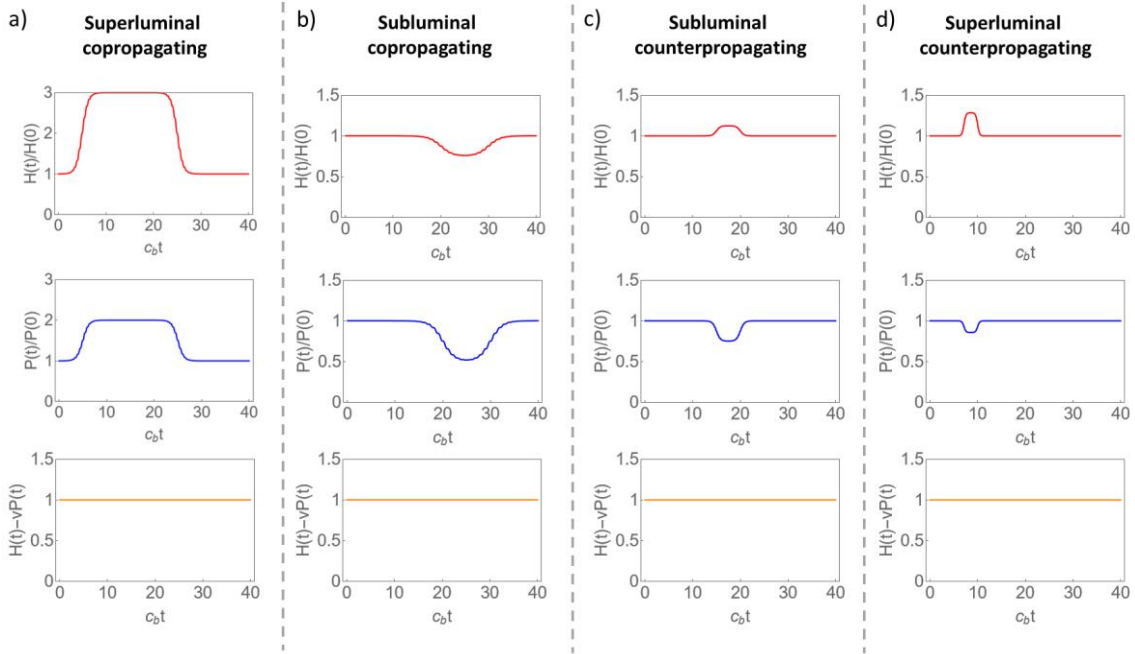

**Fig. 4. Time evolution of the energy, momentum and energy-momentum.** Normalized energy $H(t)/H(0)$, momentum $P(t)/P(0)$, and energy-momentum $[H(t) - vP(t)]/[H(0) - vP(0)]$, as a function of time, for three different regimes: (a) superluminal copropagating ($v = 2c_b$), (b) subluminal copropagating ($v = c_b/2$), (c) subluminal counterpropagating ($v = -c_b/2$) and (d) superluminal counter propagating ($v = -2c_b$). All simulation parameters are identical to those of Fig. 3. The calculations confirm that both energy and momentum change in time, while the energy-momentum remains a conserved quantity.

## 8. Conclusions

Our analysis highlights that USTMs exhibit a spatiotemporal translation symmetry with a characteristic velocity defined by the velocity of the modulation. Following Noether's theorem, the associated conserved quantity is an energy-momentum, i.e., a linear combination of the energy and Minkowski momentum, weighted by the velocity of the modulation. Moreover, the conservation of the energy-momentum imposes some constraints on the range of light-matter interactions that can take place within a USTMs. Specifically, changes of energy and momentum are fundamentally intertwined by the modulation velocity. We expect that understanding these constraints will provide better design guidelines for the engineering of different light-matter interaction regimes in the realm of USTMs, and will motivate research on more general spacetime metamaterials that overcome them.

In general, our results highlight the existence of reduced symmetries and conserved quantities in spatiotemporal metamaterials, and the need to account for them in order to fully understand the range of allowed interactions. In other words, spacetime metamaterials break both spatial and temporal symmetries, but there are more generalized symmetries that should be considered. We expect that our research will inspire further research in this direction within the context of spacetime metamaterials. Finally, our results also broadly highlight the role of Noether's theorem in understanding light-matter interactions in photonics, which might foster its applicability in related fields.

## Acknowledgements

This work was supported by ERC Starting Grant No. ERC-2020-STG948504-NZINATECH. J.E.V.-L. acknowledges support from Juan de la Cierva–Formación fellowship FJC2021-047776-I. I.L. further acknowledges support from Ramón y Cajal fellowship RYC2018-024123-I.

## References

(1) Zhou, Y.; Alam, M. Z.; Karimi, M.; Upham, J.; Reshef, O.; Liu, C.; Willner, A. E.; Boyd, R. W. Broadband Frequency Translation through Time Refraction in an Epsilon-near-Zero Material. *Nat. Commun.* **2020**, *11* (1), 2180. https://doi.org/10.1038/s41467-020-15682-2.
(2) Liu, C.; Alam, M. Z.; Pang, K.; Manukyan, K.; Reshef, O.; Zhou, Y.; Choudhary, S.; Patrow, J.; Pennathurs, A.; Song, H.; Zhao, Z.; Zhang, R.; Alishahi, F.; Fallahpour, A.; Cao, Y.; Almaiman, A.; Dawlaty, J. M.; Tur, M.; Boyd, R. W.; Willner, A. E. Photon Acceleration Using a Time-Varying Epsilon-near-Zero Metasurface. *ACS Photonics* **2021**, *8* (3), 716–720. https://doi.org/10.1021/acsphotonics.0c01929.
(3) Tirole, R.; Vezzoli, S.; Galiffi, E.; Robertson, I.; Maurice, D.; Tilmann, B.; Maier, S. A.; Pendry, J. B.; Sapienza, R. Double-Slit Time Diffraction at Optical Frequencies. *Nat. Phys.* **2023**, *19* (7), 999–1002. https://doi.org/10.1038/s41567-023-01993-w.
(4) Tirole, R.; Galiffi, E.; Dranczewski, J.; Attavar, T.; Tilmann, B.; Wang, Y.-T.; Huidobro, P. A.; Alù, A.; Pendry, J. B.; Maier, S. A.; Vezzoli, S.; Sapienza, R. Saturable Time-Varying Mirror Based on an Epsilon-Near-Zero Material. *Phys. Rev. Appl.* **2022**, *18* (5), 054067. https://doi.org/10.1103/PhysRevApplied.18.054067.
(5) Lustig, E.; Segal, O.; Saha, S.; Bordo, E.; Chowdhury, S. N.; Sharabi, Y.; Fleischer, A.; Boltasseva, A.; Cohen, O.; Shalaev, V. M.; Segev, M. Time-Refraction Optics with Single Cycle Modulation. *Nanophotonics* **2023**, *12* (12), 2221–2230. https://doi.org/10.1515/nanoph-2023-0126.
(6) Harwood, A. C.; Vezzoli, S.; Raziman, T. V.; Hooper, C.; Tirole, R.; Wu, F.; Maier, S.; Pendry, J. B.; Horsley, S. A. R.; Sapienza, R. Super-Luminal Synthetic Motion with a Space-Time Optical Metasurface. arXiv July 15, 2024. https://doi.org/10.48550/arXiv.2407.10809.
(7) Moussa, H.; Xu, G.; Yin, S.; Galiffi, E.; Ra'di, Y.; Alù, A. Observation of Temporal Reflection and Broadband Frequency Translation at Photonic Time Interfaces. *Nat. Phys.* **2023**, *19* (6), 863–868. https://doi.org/10.1038/s41567-023-01975-y.
(8) Reyes-Ayona, J. R.; Halevi, P. Observation of Genuine Wave Vector (k or β) Gap in a Dynamic Transmission Line and Temporal Photonic Crystals. *Appl. Phys. Lett.* **2015**, *107* (7), 074101. https://doi.org/10.1063/1.4928659.
(9) *Revealing non-Hermitian band structure of photonic Floquet media | Science Advances*. https://www.science.org/doi/10.1126/sciadv.abo6220 (accessed 2024-06-20).
(10) Metasurface-Based Realization of Photonic Time Crystals. https://www.science.org/doi/full/10.1126/sciadv.adg7541 (accessed 2024-06-20).
(11) Engheta, N. Four-Dimensional Optics Using Time-Varying Metamaterials. *Science* **2023**, *379* (6638), 1190–1191. https://doi.org/10.1126/science.adf1094.
(12) Galiffi, E.; Tirole, R.; Yin, S.; Li, H.; Vezzoli, S.; Huidobro, P. A.; Silveirinha, M. G.; Sapienza, R.; Alù, A.; Pendry, J. B. Photonics of Time-Varying Media. *Adv. Photonics* **2022**, *4* (1), 014002. https://doi.org/10.1117/1.AP.4.1.014002.
(13) Caloz, C.; Deck-Léger, Z.-L. Spacetime Metamaterials—Part II: Theory and Applications. *IEEE Trans. Antennas Propag.* **2020**, *68* (3), 1583–1598. https://doi.org/10.1109/TAP.2019.2944216.
(14) Caloz, C.; Deck-Léger, Z.-L.; Bahrami, A.; Vicente, O. C.; Li, Z. Generalized Space-Time Engineered Modulation (GSTEM) Metamaterials: A Global and Extended Perspective. *IEEE Antennas Propag. Mag.* **2023**, *65* (4), 50–60. https://doi.org/10.1109/MAP.2022.3216773.
(15) Kosmann-Schwarzbach, Y. The Noether Theorems. In *The Noether Theorems: Invariance and Conservation Laws in the Twentieth Century*; Kosmann-Schwarzbach, Y., Schwarzbach, B. E., Eds.; Springer: New York, NY, 2011; pp 55–64. https://doi.org/10.1007/978-0-387-87868-3_3.
(16) Cohen-Tannoudji, C.; Dupont-Roc, J.; Grynberg, G. *Photons and Atoms: Introduction to Quantum Electrodynamics*; Physics textbook; Wiley-VCH: Weinheim, 2004.
(17) Vázquez-Lozano, J. E.; Liberal, I. Shaping the Quantum Vacuum with Anisotropic Temporal Boundaries. *Nanophotonics* **2023**, *12* (3), 539–548. https://doi.org/10.1515/nanoph-2022-0491.
(18) Ganfornina-Andrades, A.; Vázquez-Lozano, J. E.; Liberal, I. Quantum Vacuum Amplification in Time-Varying Media with Arbitrary Temporal Profiles. arXiv December 20, 2023. http://arxiv.org/abs/2312.13315 (accessed 2024-06-21).
(19) Vázquez-Lozano, J. E.; Liberal, I. Incandescent Temporal Metamaterials. arXiv October 11, 2022. https://doi.org/10.48550/arXiv.2210.05565.
(20) Liberal, I.; Vázquez-Lozano, J. E.; Pacheco-Peña, V. Quantum Antireflection Temporal Coatings: Quantum State Frequency Shifting and Inhibited Thermal Noise Amplification. *Laser Photonics Rev.* **2023**, *17* (9), 2200720. https://doi.org/10.1002/lpor.202200720.

(21) Lyubarov, M.; Lumer, Y.; Dikopoltsev, A.; Lustig, E.; Sharabi, Y.; Segev, M. Amplified Emission and Lasing in Photonic Time Crystals. *Science* **2022**, *377* (6604), 425–428. https://doi.org/10.1126/science.abo3324.
(22) *Light emission by free electrons in photonic time-crystals | PNAS*. https://www.pnas.org/doi/abs/10.1073/pnas.2119705119 (accessed 2024-06-20).
(23) Sounas, D. L.; Alù, A. Non-Reciprocal Photonics Based on Time Modulation. *Nat. Photonics* **2017**, *11* (12), 774–783. https://doi.org/10.1038/s41566-017-0051-x.
(24) Barnett, S. M. Resolution of the Abraham-Minkowski Dilemma. *Phys. Rev. Lett.* **2010**, *104* (7), 070401. https://doi.org/10.1103/PhysRevLett.104.070401.
(25) Lobet, M.; Liberal, I.; Vertchenko, L.; Lavrinenko, A. V.; Engheta, N.; Mazur, E. Momentum Considerations inside Near-Zero Index Materials. *Light Sci. Appl.* **2022**, *11* (1), 110. https://doi.org/10.1038/s41377-022-00790-z.
(26) Ortega-Gomez, A.; Lobet, M.; Vázquez-Lozano, J. E.; Liberal, I. Tutorial on the Conservation of Momentum in Photonic Time-Varying Media [Invited]. *Opt. Mater. Express* **2023**, *13* (6), 1598–1608. https://doi.org/10.1364/OME.485540.
(27) Jajin, M. M.; Vázquez-Lozano, J. E.; Liberal, I. Symmetries and Conservation of Spin Angular Momentum, Helicity, and Chirality in Photonic Time-Varying Media. arXiv April 18, 2024. http://arxiv.org/abs/2404.12340 (accessed 2024-06-20).
(28) Galiffi, E.; Huidobro, P. A.; Pendry, J. B. Broadband Nonreciprocal Amplification in Luminal Metamaterials. *Phys. Rev. Lett.* **2019**, *123* (20), 206101. https://doi.org/10.1103/PhysRevLett.123.206101.
(29) Pendry, J. B.; Galiffi, E.; Huidobro, P. A. Gain in Time-Dependent Media—a New Mechanism. *JOSA B* **2021**, *38* (11), 3360–3366. https://doi.org/10.1364/JOSAB.427682.
(30) Scarborough, C.; Grbic, A. Generalized Eigenvalue Problem for Spatially Discrete Traveling-Wave-Modulated Circuit Networks. *IEEE Trans. Microw. Theory Tech.* **2023**, *71* (2), 511–521. https://doi.org/10.1109/TMTT.2022.3225321.
(31) Taravati, S.; Eleftheriades, G. V. Generalized Space-Time-Periodic Diffraction Gratings: Theory and Applications. *Phys. Rev. Appl.* **2019**, *12* (2), 024026. https://doi.org/10.1103/PhysRevApplied.12.024026.
(32) Taravati, S.; Eleftheriades, G. V. Microwave Space-Time-Modulated Metasurfaces. *ACS Photonics* **2022**, *9* (2), 305–318. https://doi.org/10.1021/acsphotonics.1c01041.
(33) Wu, Z.; Scarborough, C.; Grbic, A. Space-Time-Modulated Metasurfaces with Spatial Discretization: Free-Space $N$-Path Systems. *Phys. Rev. Appl.* **2020**, *14* (6), 064060. https://doi.org/10.1103/PhysRevApplied.14.064060.
(34) Moreno-Rodríguez, S.; Alex-Amor, A.; Padilla, P.; Valenzuela-Valdés, J. F.; Molero, C. Space-Time Metallic Metasurfaces for Frequency Conversion and Beamforming. *Phys. Rev. Appl.* **2024**, *21* (6), 064018. https://doi.org/10.1103/PhysRevApplied.21.064018.
(35) Horsley, S. A. R.; Pendry, J. B. Quantum Electrodynamics of Time-Varying Gratings. *Proc. Natl. Acad. Sci.* **2023**, *120* (36), e2302652120. https://doi.org/10.1073/pnas.2302652120.
(36) Pendry, J. B.; Horsley, S. A. R. QED in Space–Time Varying Materials. *APL Quantum* **2024**, *1* (2), 020901. https://doi.org/10.1063/5.0199503.
(37) Oue, D.; Ding, K.; Pendry, J. B. Čerenkov Radiation in Vacuum from a Superluminal Grating. *Phys. Rev. Res.* **2022**, *4* (1), 013064. https://doi.org/10.1103/PhysRevResearch.4.013064.
(38) Oue, D.; Ding, K.; Pendry, J. B. Noncontact Frictional Force between Surfaces by Peristaltic Permittivity Modulation. *Phys. Rev. A* **2023**, *107* (6), 063501. https://doi.org/10.1103/PhysRevA.107.063501.
(39) Huidobro, P. A.; Galiffi, E.; Guenneau, S.; Craster, R. V.; Pendry, J. B. Fresnel Drag in Space–Time-Modulated Metamaterials. *Proc. Natl. Acad. Sci.* **2019**, *116* (50), 24943–24948. https://doi.org/10.1073/pnas.1915027116.
(40) Bahrami, A.; Deck-Léger, Z.-L.; Caloz, C. Electrodynamics of Accelerated-Modulation Space-Time Metamaterials. *Phys. Rev. Appl.* **2023**, *19* (5), 054044. https://doi.org/10.1103/PhysRevApplied.19.054044.
(41) Deck-Léger, Z.-L.; Bahrami, A.; Li, Z.; Caloz, C. Generalized FDTD Scheme for the Simulation of Electromagnetic Scattering in Moving Structures. *Opt. Express* **2023**, *31* (14), 23214–23228. https://doi.org/10.1364/OE.493099.
(42) Kong, J. A. *Theory of Electromagnetic Waves*; A Wiley-Interscience publication; Wiley: New York, 1975.
(43) Ziolkowski, R. W.; Arnold, J. M.; Gogny, D. M. Ultrafast Pulse Interactions with Two-Level Atoms. *Phys. Rev. A* **1995**, *52* (4), 3082–3094. https://doi.org/10.1103/PhysRevA.52.3082.
(44) Pendry, J. B. Air Conditioning for Photons [Invited]. *Opt. Mater. Express* **2024**, *14* (2), 407–413. https://doi.org/10.1364/OME.511182.
(45) John, F. *Partial Differential Equations*, 4th ed.; Applied mathematical sciences; Springer-Verlag: New York, 1982.